\documentclass[12pt]{article}
\usepackage{epsfig}
\def\be{\begin{equation}}
\def\ee{\end{equation}}
\def\bea{\begin{eqnarray}}
\def\eea{\end{eqnarray}}
\usepackage{graphicx}

\catcode`\@=11
\def\lsim{\mathrel{\mathpalette\@versim<}}
\def\gsim{\mathrel{\mathpalette\@versim>}}
\def\@versim#1#2{\vcenter{\offinterlineskip
\ialign{$\m@th#1\hfil##\hfil$\crcr#2\crcr\sim\crcr } }}
\catcode`\@=12
\usepackage{axodraw}

\catcode`@=12
\begin{document}
\thispagestyle{empty}
\begin{flushright}
UCRHEP-T530\\ZTF-EP-13-03\\
August 2013\
\end{flushright}
\vspace{0.6in}
\begin{center}
{\LARGE \bf New Scotogenic Model of Neutrino Mass\\
with $U(1)_D$ Gauge Interaction\\}
\vspace{1.2in}
{\bf Ernest Ma$^1$, Ivica Picek$^2$, and Branimir Radov\v ci\' c$^2$\\}
\vspace{0.2in}
{\sl $^1$ Department of Physics and Astronomy, University of California,\\
Riverside, California 92521, USA\\}
\vspace{0.1in}
{\sl $^2$ Department of Physics, Faculty of Science, University of Zagreb,\\
P.O.B. 331, HR-10002 Zagreb, Croatia\\}
\end{center}
\vspace{1.2in}
\begin{abstract}\
We propose a new realization of the one-loop radiative model of neutrino mass
generated by dark matter (scotogenic), where the particles in the loop have an
additional $U(1)_D$ gauge symmetry, which may be exact or broken to $Z_2$.
This model is relevant to a number of astrophysical observations, including
AMS-02 and the dark matter distribution in dwarf galactic halos.
\end{abstract}

\newpage
\baselineskip 24pt

\begin{figure}[htb]
\begin{center}
\begin{picture}(360,120)(0,0)
\ArrowLine(90,10)(130,10)
\ArrowLine(180,10)(130,10)
\ArrowLine(180,10)(230,10)
\ArrowLine(270,10)(230,10)
\DashArrowLine(155,85)(180,60)3
\DashArrowLine(205,85)(180,60)3
\DashArrowArc(180,10)(50,90,180)3
\DashArrowArcn(180,10)(50,90,0)3

\Text(110,0)[]{\large $\nu_i$}
\Text(250,0)[]{\large $\nu_j$}
\Text(180,0)[]{\large $N_k$}
\Text(135,50)[]{\large $\eta^0$}
\Text(230,50)[]{\large $\eta^0$}
\Text(150,90)[]{\large $\phi^{0}$}
\Text(217,90)[]{\large $\phi^{0}$}

\end{picture}
\end{center}
\caption{One-loop generation of neutrino mass with $Z_2$ symmetry.}
\end{figure}
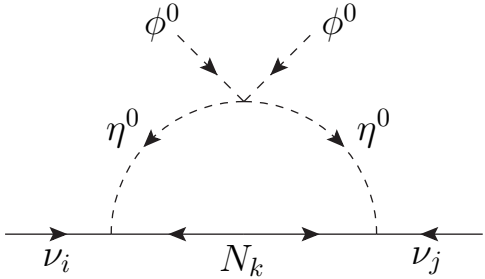

The notion that dark matter (DM) is the origin of neutrino mass (scotogenic)
is by now a common theme among many studies.  The first one-loop
realization~\cite{Ma:2006km}, as shown in Fig.~1, remains the simplest such
example.  The standard model (SM) of quark and lepton interactions is
augmented by three neutral singlet Majorana fermions $N_{1,2,3}$ and a second
scalar doublet $(\eta^+,\eta^0)$.  A new discrete $Z_2$ symmetry is imposed so
that the new particles are odd and all the SM particles even.  The complex
scalar $\eta^0 = (\eta_R + i \eta_I)/\sqrt{2}$ is split by the allowed
$(\lambda_5/2)(\Phi^\dagger \eta)^2 + H.c.$ term in the Higgs potential
so that $m_R \neq m_I$ and the scotogenic neutrino mass is given
by~\cite{Ma:2006km}
\begin{equation}
({\cal M}_\nu)_{ij} = \sum_k {h_{ik} h_{jk} M_k \over 16 \pi^2}
\left[ {m_R^2 \over m_R^2 - M_k^2} \ln {m_R^2 \over M_k^2} -
{m_I^2 \over m_I^2 - M_k^2} \ln {m_I^2 \over M_k^2} \right].
\end{equation}
The DM candidate is either $\eta_R$ (assuming of course that $m_R < m_I$) or
$N_1$ (assuming of course $M_1 < M_{2,3}$).  Many studies and variations
of this original model are now available in the literature.  One important
extension is the promotion of the stabilizing discrete $Z_2$ symmetry to
a $U(1)_D$ gauge symmetry~\cite{Kubo:2006rm,Ma:2008ba}, which gets broken
to $Z_2$ through an additional scalar field.  This has two effects: (1) the
stability of dark matter is now protected against possible violation of the
$Z_2$ symmetry from higher-dimensional operators including those of quantum
gravity, (2) the force carriers (both vector and
scalar) between DM particles may be relevant in explaining a number of
astrophysical observations.

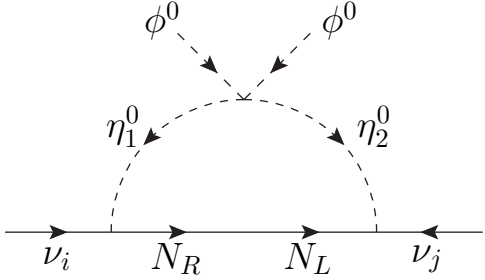
\begin{figure}[htb]
\begin{center}
\begin{picture}(360,120)(0,0)
\ArrowLine(90,10)(130,10)
\ArrowLine(130,10)(180,10)
\ArrowLine(180,10)(230,10)
\ArrowLine(270,10)(230,10)
\DashArrowLine(155,85)(180,60)3
\DashArrowLine(205,85)(180,60)3
\DashArrowArc(180,10)(50,90,180)3
\DashArrowArcn(180,10)(50,90,0)3

\Text(110,0)[]{\large $\nu_i$}
\Text(250,0)[]{\large $\nu_j$}
\Text(155,0)[]{\large $N_R$}
\Text(205,0)[]{\large $N_L$}
\Text(135,50)[]{\large $\eta_1^0$}
\Text(230,50)[]{\large $\eta_2^0$}
\Text(150,90)[]{\large $\phi^{0}$}
\Text(217,90)[]{\large $\phi^{0}$}

\end{picture}
\end{center}
\caption{One-loop generation of neutrino mass with $U(1)_D$ symmetry.}
\end{figure}

In this paper, we propose a new scotogenic model with a $U(1)_D$ gauge
symmetry which may be exact or broken to $Z_2$.  The new particles are
two scalar doublets $(\eta_1^+,\eta_1^0) \sim 1$ and $(\eta_2^+,\eta_2^0)
\sim -1$ under $U(1)_D$, and three neutral singlet Dirac fermions
$N_{1,2,3} \sim 1$ under $U(1)_D$.  The allowed couplings completing
the loop, as shown in Fig.~2, are $h_1 \bar{N}_R \nu_L \eta_1^0$,
$h_2 N_L \nu_L \eta_2^0$, and $(\Phi^\dagger \eta_1)(\Phi^\dagger \eta_2)$
which mixes $\eta_1^0$ and $\bar{\eta}_2^0$.  Let
\begin{equation}
\pmatrix{\eta_1^0 \cr \bar{\eta}_2^0} = \pmatrix{\cos \theta & \sin \theta \cr
-\sin \theta & \cos \theta} \pmatrix{\chi_1 \cr \chi_2},
\end{equation}
where $\chi_{1,2}$ are mass eigenstates, then the analog of Eq.~(1) becomes
\begin{equation}
({\cal M}_\nu)_{ij} = \sin \theta \cos \theta \sum_k {[(h_1)_{ki} (h_2)_{kj} +
(h_2)_{ki} (h_1)_{kj}] M_k \over 8 \pi^2}
\left[ {m_1^2 \over m_1^2 - M_k^2} \ln {m_1^2 \over M_k^2} -
{m_2^2 \over m_2^2 - M_k^2} \ln {m_2^2 \over M_k^2} \right],
\end{equation}
where $m_{1,2}$ are the masses of $\chi_{1,2}$ and $M_k$ the mass of $N_k$.
Note that in contrast to Fig.~1, Majorana neutrino masses are obtained in
Fig.~2 even though only Dirac masses appear in the loop.
At this point, the $U(1)_D$ gauge symmetry may remain exact, in
which case there is a massless dark photon. However,
we can also break the $U(1)_D$ gauge symmetry to $Z_2$ by a complex singlet
scalar field $\zeta \sim 2$, in which case there is a massive dark
photon $\gamma_D$ as well as a dark Higgs boson, both of which may be
relevant in astrophysics as force carriers between DM particles.

If $U(1)_D$ is unbroken, only $N_1$ is a DM candidate because $\eta^0_{1,2}$ are
not split in their real and imaginary parts, which means that their
interaction with nuclei through $Z$ exchange cannot be suppressed
and thus ruled out by direct-search data as a possible DM candidate.
In the presence of $U(1)_D$ breaking with the allowed $y_L \zeta^\dagger
N_L N_L$ and $y_R \zeta^\dagger N_R N_R$ couplings, $N$ is no longer a Dirac
fermion, but if these new terms
are small, it may still be a pseudo-Dirac particle.  At the same time, the
$\zeta \eta_1^\dagger \eta_2$ coupling allows splitting of the real and
imaginary parts of $\eta_{1,2}$.

There is yet another scenario, where the gauge $U(1)_D$ symmetry becomes
an exact global $U(1)_D$ symmetry.  This is accomplished if $\zeta$ is
forbidden to couple to $N$ or $\eta_{1,2}$, by choosing for example
$\zeta \sim 3$.  The spontaneous breaking of the gauge $U(1)_D$ symmetry
now results in a global $U(1)_D$ symmetry, under which only $N$ and
$\eta_{1,2}$ transform.  This means that dark Higgs is no longer a
force carrier for the dark matter $N$, but the vector force carrier
$\gamma_D$ remains and is no longer massless.

In the following we choose our DM candidate to be the lightest Dirac
(or pseudo-Dirac) $N$ and investigate how it fits into the standard thermal
WIMP (Weakly Interacting Massive Particle) paradigm. The dark photon
$\gamma_D$ may be
massless~\cite{Ackerman:2008gi} in which case a realistic scenario
would require $N$ to be heavier than about 1 TeV.  If $U(1)_D$ is broken by
$\zeta = (u + \rho + i \sigma)/\sqrt{2}$, where $u = \sqrt{2} \langle \zeta
\rangle$, then $\gamma_D$ is massive together with $\rho$.  In the following
we will assume $u$ to be small compared to the decoupling temperature of $N$,
in which case its relic abundance is determined by the unbroken theory,
whereas at present, its interaction with ordinary matter is determined by
the broken theory.   In the early Universe, $N \bar{N}$ would
annihilate to $\gamma_D \gamma_D$ and $\zeta \zeta^*$. Since the dark
scalar singlet $\zeta$ must mix with the SM Higgs doublet $\Phi$
in the most general scalar potential containing both, and the dark photon
$\gamma_D$ may have kinetic mixing~\cite{Holdom:1985ag} with the SM photon,
these processes will allow $N$ to have the correct thermal relic abundance
to be a suitable DM candidate.  Furthermore, for $\gamma_D$ and $\rho$
lighter than about 0.1 GeV, a number of astrophysical observations at
present may be explained.

Our DM scenario assumes $N$ to be much heavier than the $U(1)_D$ breaking
scale.  Thus $N$ is in general pseudo-Dirac.  As far as relic abundance
is concerned, it behaves as a Dirac fermion~\cite{DeSimone:2010tf}.
Further, since it can annihilate into scalars $(\zeta \zeta^*)$ or
vectors $(\gamma_D \gamma_D)$ instead of just SM quarks and leptons, its
cross section is not suppressed by fermion mass. Its thermally averaged
s-wave annihilation cross sections to $\gamma_D \gamma_D$ and $\zeta \zeta^*$
are given by
\begin{equation}
\langle\sigma (N \bar{N} \to \gamma_D \gamma_D) v\rangle = \frac{\pi
\alpha^2_D}{M_1^2}\ ,
\end{equation}
\begin{equation}
\langle\sigma (N \bar{N} \to \zeta \zeta^*) v\rangle =
\frac{(|y_L|^2+|y_R|^2)^2 -
(y_L y_R^* - y_L^* y_R)^2}{16 \pi M_1^2} \ ,
\end{equation}
where $\alpha_D = g_D^2/4 \pi$ is the dark fine structure constant and
we have neglected the masses of $\gamma_D$ and $\zeta$.

\begin{figure}[h]
\centerline{\includegraphics[scale=0.70]{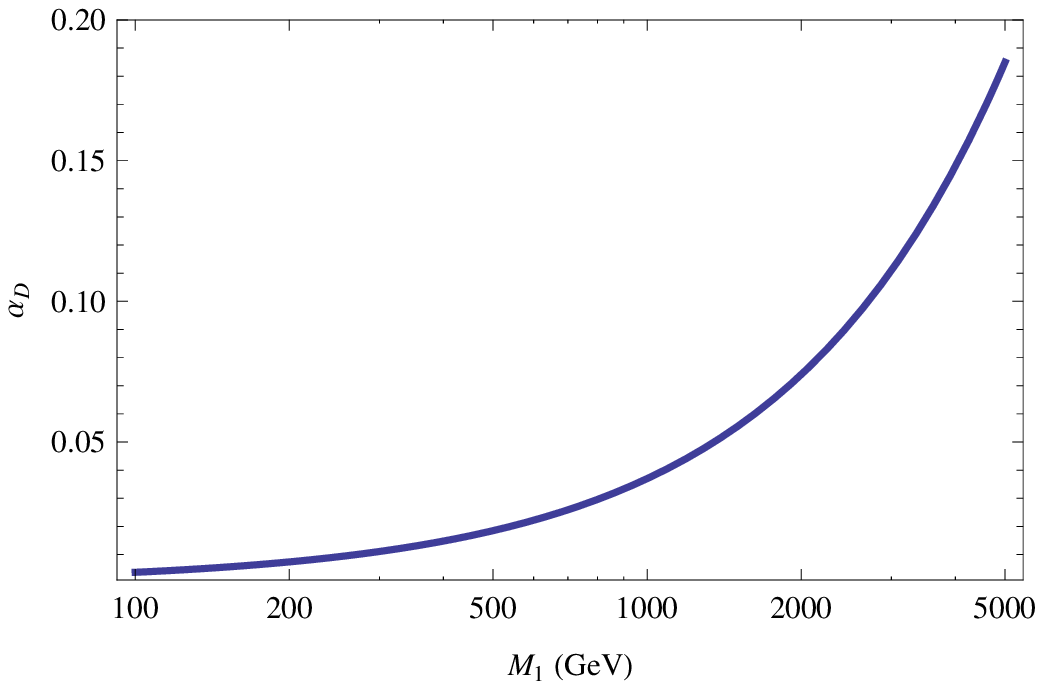}\hspace{1cm}
\includegraphics[scale=0.70]{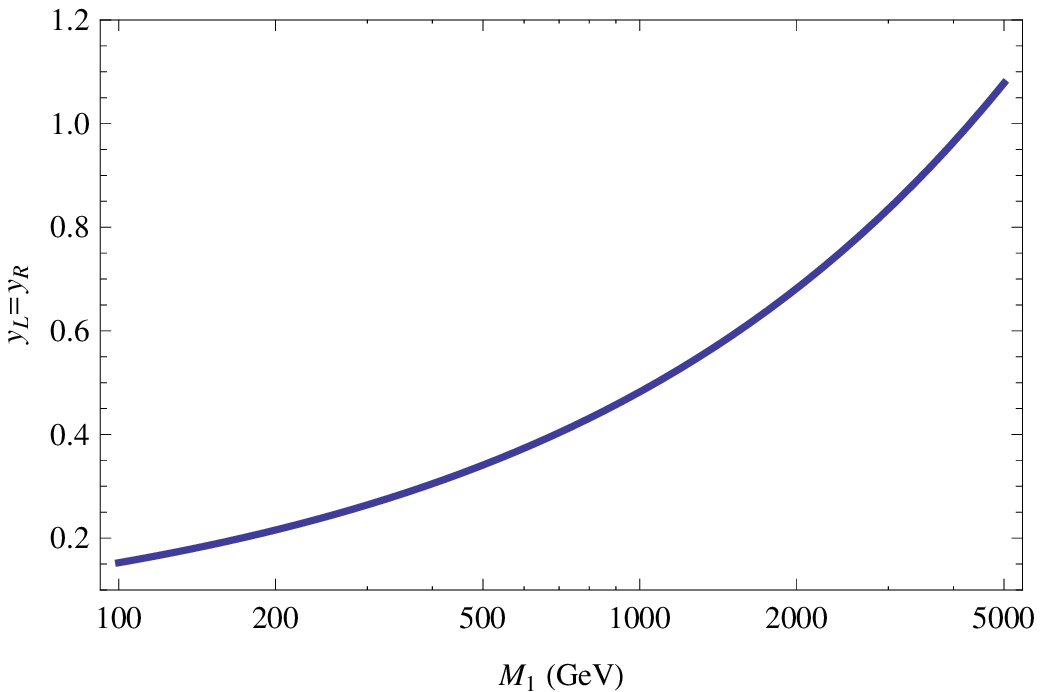}}
\caption{\small Values of DM couplings $\alpha_D$ (left) and $y_L,y_R$ (right)
as a function od DM mass required to obtain observed relic abundance of DM
in the Universe. For simplicity we have chosen $y_L=y_R$. }
\end{figure}

In Fig.~3 we display the values of DM couplings required to obtain the
observed value for the dark-matter relic density of the Universe,
$\Omega_{DM} h^2 = 0.1187(17)$~\cite{Ade:2013zuv}.
For example, if $M_1=1$ TeV, then we need either $\alpha_D=0.04$
or $y_L=y_R=0.48$.

As $U(1)_D$ is broken, the Dirac DM fermion $N$ splits up into two Majorana
fermions of about equal mass,  The heavier state $\Sigma_2$ will decay into
the lighter state $\Sigma_1$ and a force carrier ($\Sigma_2 \to \Sigma_1
\gamma_D , \Sigma_1 \rho$) if kinematically allowed.
If the mass splitting is smaller than the mass of the force carriers,
$\Sigma_2$ will decay through an off-shell force carrier or $\eta_{1,2}$
to $\Sigma_1$ and a pair of SM leptons.

There are two important phenomenological implications of our $U(1)_D$
DM scenario.  First, the large positron excess observed by
PAMELA~\cite{Adriani:2008zr,Adriani:2010ib} requires an enhancement of
the DM annihilation cross section at present compared to what it was
at the time of freeze-out.  This may be accomplished~\cite{ArkaniHamed:2008qn}
by the inclusion of a new force in the dark sector, resulting in
a Sommerfeld enhancement of the cross section from multiple exchange
of the light force carrier. Recent AMS-02 results~\cite{Aguilar:2013qda}
may also be explained~\cite{Liu:2013vha} in a similar way.
In our case, since $\rho$ mixes with $h$ and $\gamma_D$ mixes with $\gamma$,
their decays to $\mu^- \mu^+$ and $e^- e^+$ are ideal for such a purpose.

Second, DM self interactions change its density profile from the usual
collisionless WIMP scenario.  To reconcile the theoretical prediction
with the present astronomical observation of the halos of dwarf galaxies,
a rather large cross section per unit DM mass $\sim 1 ~{\rm cm}^2/{\rm g}$ is
required, and may be achieved~\cite{Tulin:2013teo,Kaplinghat:2013kqa}
with rather light force mediators, such as $M_1 = 1$ TeV and
$m_{\rho,\gamma_D} \sim 4$ MeV, or $M_1 = 100$ GeV and $m_{\rho,\gamma_D} \sim 20$ MeV.

Finally, additional insight into DM candidates in our scenario may come
from direct detection experiments. The current XENON100
limits~\cite{Aprile:2012nq}
are already sensitive to very small couplings corresponding to the
mixing of the dark-force carriers
with the appropriate SM bosons. For a benchmark value $10^{-10}$ for the
coupling involved in the kinetic mixing of the dark photon with the
SM photon, and for a 10-100 MeV dark force-carrier mass, XENON100 excludes
self-interacting DM with a mass larger than
$\sim 300$ GeV~\cite{Kaplinghat:2013kqa}.\\

\noindent \underline{Acknowledgment}~:~The work of EM is supported in part
by the U.~S.~Department of Energy under Grant No.~DE-FG03-94ER40837.
IP and BR are supported by the Croatian Ministry  of Science, Education and
Sports under Contract No. 119-0982930-1016.

\bibliographystyle{unsrt}

\end{document}